\documentclass[3p,onecolumn, twoside]{elsarticle}

\usepackage{amsmath}
\usepackage{nicefrac,xfrac}

\usepackage{multirow}
\usepackage{siunitx}
\usepackage{subcaption}
\usepackage{tabularx}
\journal{Nuclear Instruments and Methods}
\DeclareSIUnit\angstrom{\text{\AA}}
\newcommand{\pendo}{pendell\"osung } 
\newcommand{\Pendo}{Pendell\"osung } 
\newcommand{\sample}{$\mathrm{Ni}_{60}\mathrm{Cu}_{40}$}
\newcommand{\NCNR}{NIST Center for Neutron Research (NCNR)\renewcommand{\NCNR}{NCNR}}
\newcommand{\NIST}{National Institute of Standards and Technology (NIST)\renewcommand{\NIST}{NIST}}
\newcommand{\NIOFh}{Neutron Interferometry and Optics Facility (\mbox{NIOF-\textbf{h}})\renewcommand{\NIOFh}{\mbox{NIOF-\textbf{h}}}}
\newcommand{\NIOFa}{Neutron Interferometry and Optics Facility-a (\mbox{NIOF-\textbf{a}})\renewcommand{\NIOFa}{\mbox{NIOF-\textbf{a}}}}

\begin{document}
\begin{frontmatter}

\title{Environmental Stabilization of Perfect-Crystal Neutron Interferometry Using a Large Vacuum Chamber with Cryogenic Sample Access}

\author[NCState]{R. Valdillez}
\author[IQC,UW2]{D. G. Cory}
\author[Tulane]{R. W. Haun}
\author[NIST]{B. Heacock}
\author[NIST]{C. Heikes}
\author[NIST]{S. F. Hoogerheide}
\author[NIST]{M. G. Huber*}
\ead{michael.huber@nist.gov}
\cortext[cor1]{Corresponding author}
\author[IQC,UW1]{T. Mineeva}
\author[NIST]{J. Paster}
\author[UB]{D. Sarenac}
\author[IQC,UW1]{D. A. Pushin}
\author[NCState]{A. R. Young}

\address[NCState]{Department of Physics, North Carolina State University, Raleigh, NC 27695, USA}
\address[IQC]{Institute for Quantum Computing, University of Waterloo, Waterloo, ON, Canada, N2L3G1}
\address[UW2]{Department of Chemistry, University of Waterloo,Waterloo, ON, Canada N2L 3G1}
\address[Tulane]{Physics and Engineering Physics Dept., Tulane University, New Orleans, LA 70188, USA}
\address[NIST]{National Institute of Standards and Technology, Gaithersburg, MD 20899, USA}
\address[UW1]{Department of Physics, University of Waterloo,Waterloo, ON, Canada N2L 3G1}
\address[UB]{Department of Physics, University at Buffalo, Buffalo, NY, 14260, USA}

\begin{abstract}
Perfect-crystal neutron interferometry has been a useful tool in measuring nuclear-interactions, probing fundamental physics, and exploring quantum phenomenon.
Historically, neutron interferometry experiments have been carried out at room temperature and standard atmospheric pressure.
However, neutron interferometry is sensitive to changes in the local environment, especially thermal gradients across the crystal, resulting in phase drifts and systematic uncertainty. 
A need for measurements performed in different sample environments compound these issues.  
Fortunately,  the use of a vacuum chamber has been shown to be an effective method of environmental isolation for perfect-crystal neutron interferometers. 
A large volume, highly versatile vacuum chamber has been installed at the Neutron Interferometry and Optics Facility at the NIST Center for Neutron Research to isolate interferometry from local temperature and pressure deviations as well as allowing for the introduction of cryogenically cooled samples.
The prospect of incorporating a cryostat within a neutron interferometer opens up new areas of investigation, such as superconductivity. 
In addition to describing the vacuum chamber, we report on the first measurement of a cryogenic-cooled sample by a neutron interferometer. 
For this demonstration contrast was measured with a  \sample{} sample between \SI{4}{\K} to \SI{300}{\K}.
\end{abstract}

\begin{keyword}
Neutron Interferometry, Perfect Crystal
\end{keyword}
\end{frontmatter}

\section{Introduction}\label{sec:Intro}

Neutron interferometry (NI), utilizing a perfect silicon crystal, was first demonstrated in 1974 \cite{Rauch1974}. 
It has since proved to be a useful tool for studying a range of physics problems \cite{NIBook}. 
These include precision measurements of coherent and incoherent scattering lengths \cite{Kaiser_1979_Physik,Haun_2020_PRL,Schoen_2003_Phys_Rev.}, phase contrast imaging \cite{Pushin_2007_Appl.Phys.Lett.}, gravitationally induced quantum interference effects \cite{Littrell_1997_PhysicalReviewA,Heacock_2017_Phys.Rev.A,  Pushin_2009_PhysRev}, dark energy/5 fifth force force searches \cite{Li_2016_Phys.Rev.D}, and studies of neutron orbital angular momentum \cite{Clark_2015_Nature,Sarenac2016}.  
One area in which NI has found only limited use is in materials research. 
This is despite there are many neutron scattering instruments such as small-angle neutron scattering (SANS), triple-axis diffractometers, and reflectometers \cite{Langel2023,PRINCE2004} that are tailored for materials research.  
The inherent challenges facing the use of neutron interferometry for material science are that NI operates with lower statistics than widely-used neutron scattering instruments and it has a greater sensitivity to environmental noise.   
\par
However, the \NCNR\ is aiming towards a broader use for NI through the development of vacuum isolation and decoherence free subspace techniques from quantum information science. 
The capabilities, newly demonstrated in this paper, to cool and control a sample temperature over the range from \SI{4}{\K} to \SI{300}{\K} opens a variety of new research avenues that have been previously unavailable, particularly in condensed matter physics and materials science. 
The capabilities and characteristics of a large volume vacuum chamber and cryostat for NI are discussed below. 

\section{Interferometry}\label{sec:NI}

\subsection{Closed Loop Interferometry}\label{sec:NI:Cloop}

\begin{figure*}
    \centering
    \begin{subfigure}[b]{0.475\textwidth}
        \centering
        \includegraphics[width=0.9\columnwidth]{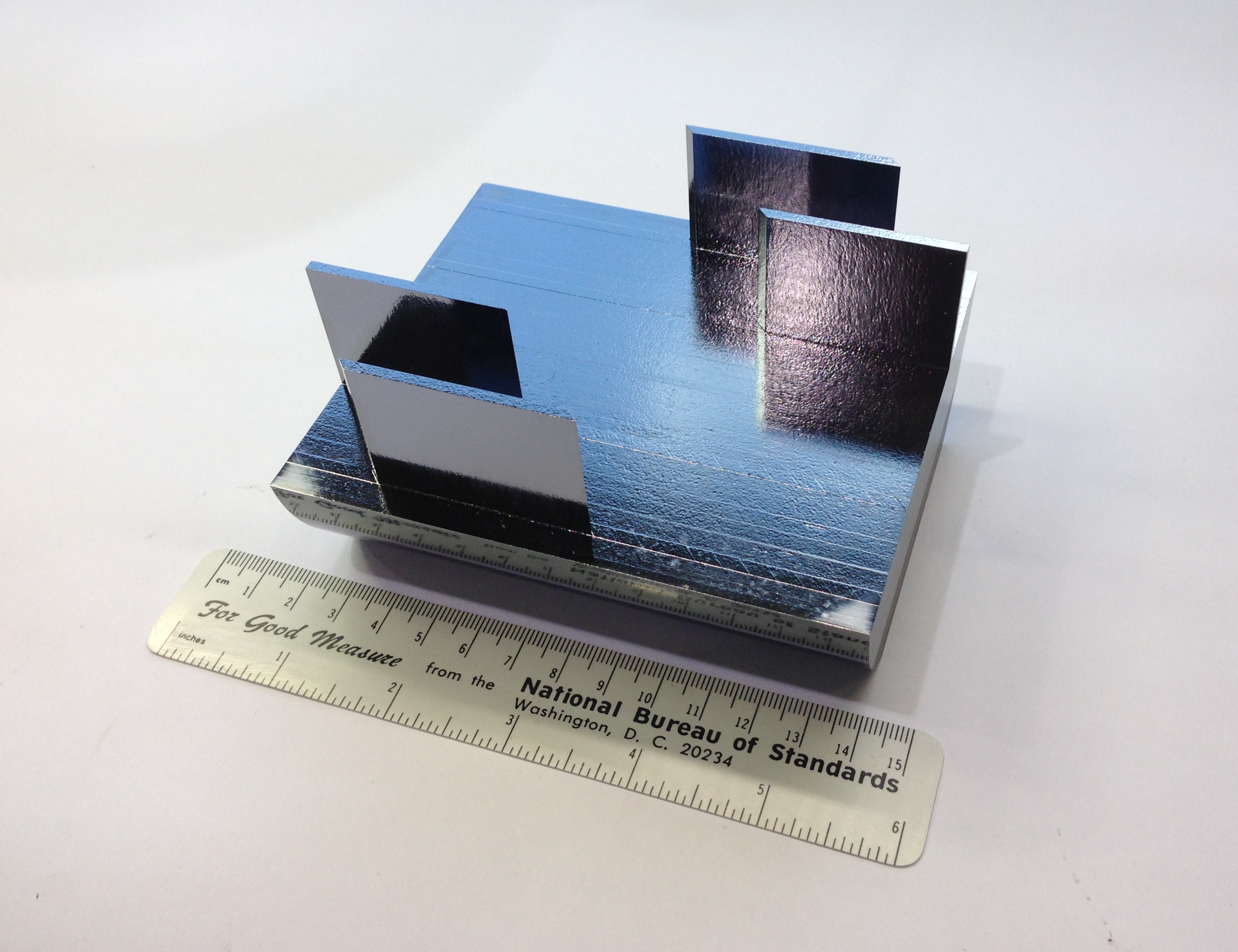}
    \end{subfigure}
    \hfill
    \begin{subfigure}[b]{0.475\textwidth}  
        \centering 
        \includegraphics[width=0.9\columnwidth]{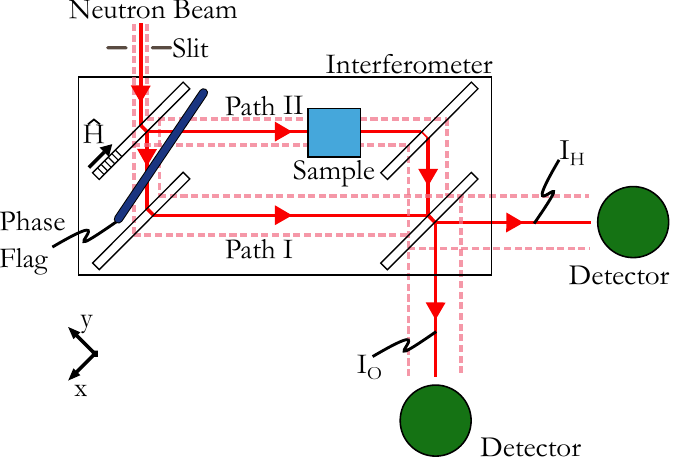}
    \end{subfigure}
    \caption[]
    {(LEFT) A photo of a skew-symmetric perfect-crystal silicon neutron interferometer. (RIGHT) A perfect-crystal neutron interferometer separates a neutron along two paths which interfere before exiting the interferometer. Changes in the relative phase shift between Path I and II modulate the intensities ($I_O$ and $I_H$) measured by the detectors. A sample called a phase flag is used to controllably vary the phase.} 
    \label{fig:ni}
\end{figure*}

Perfect silicon crystal neutron interferometry is analogous to Mach-Zehnder (MZ) interferometry for light. 
Instead of using full and half mirrors to split and recombine the incident radiation beam, perfect-crystal neutron interferometry uses Bragg diffraction to coherently split the wave into spatially separate paths. 
A neutron interferometer consists of a single perfect silicon crystal ingot machined so that there are two or more crystal blades extruding from a large common base.  
The common base provides stability and ensures that the crystal blade's atomic planes are aligned relative to one another. 
Neutrons are Bragg diffracted in the first crystal blade, acting like a beam splitter.  Subsequent blades are used to interfere the two paths as the neutron exits the interferometer.
A skew-symmetric perfect-crystal neutron interferometer, where the diffracting blades are not all equally spaced, is shown in Fig.\ \ref{fig:ni}. 
\par
Neutrons exit the last blade in two directions, one in the same direction as the incident beam (O-beam) and one in the diffracted direction (H-beam). 
Detection of the neutrons after exiting the interferometer is done using \textsuperscript{3}He proportional counters with nearly \SI{100}{\%} detection efficiency. 
The intensity of the O- and H-beams, 
\begin{equation}
\begin{split}
    I_0 &= c_0 + c_1\cos(\Delta\phi) \\ 
    I_H &= c_2 - c_1\cos(\Delta\phi),\label{eqn:intensity}
\end{split}
\end{equation}
are \SI{180}{\degree} out of phase, and since there is almost no neutron absorption in  silicon, the sum of the O- and H- beam intensities is a constant.  
In Eqn.\ \ref{eqn:intensity}, the coefficients $c_{0,1,2}$ are related to the incident neutron beam intensity and initial contrast of the interferometer but are generally treated as fit parameters.
The O-beam contrast of $\mathcal{C} = c_1/ c_0$ is regarded as a coarse gauge on the quality. Contrast can also be decreased by overall noise, beam size, sample effects, and other  parameters of the overall experiment setup. 
Large phase drifts, instability, or loss of coherence decreases $\mathcal{C}$, which for the interferometers at NIST, typically start at greater than \SI{80}{\%}.
The parameter $\Delta\phi$ is the relative phase difference between the interferometer paths I and II. 
A sample placed in path II causes a relative phase shift of
\begin{eqnarray}
\Delta\phi _\mathrm{nucl}= -\lambda \cdot N(T) \cdot b_c \cdot D  \hspace{ 12pt} \textnormal{(Nuclear)}\ \label{eqn:Phase}
\end{eqnarray}
due to the neutron-nuclei interaction.  Here $\lambda$ is the neutron wavelength, $N(T)$ is the atomic density of a sample at temperature $T$,  $b_c$ is the scattering length which is an isotope-dependent quantity, and $D$ is the sample thickness. 
The phase along each path of the interferometer may be varied by rotating a control sample called a phase flag, typically a \SI{2}{mm} thick piece of quartz.    
In addition to the neutron's interaction with a nucleus, a magnetic field  along one path generates a relative phase,
\begin{eqnarray}
\Delta\phi _\mathrm{mag} = \pm \frac{\mu_\textrm{n} m_\textrm{n}\lambda D B_0  }{2\pi\hbar^2}   \hspace{ 12pt}  \textnormal{(Magnetic)}\ \label{eqn:PhaseMag}
\end{eqnarray}
where $\mu_\textrm{n} = $ \SI{-9.649}{J\per T} is the magnetic moment of the neutron, $m_\textrm{n} = $ \SI{1.675d-27}{kg} is the neutron mass, and here $D$ is the length of the region with constant magnetic field $B_0$ . 
\subsection{\Pendo Interferometry}\label{sec:NI:pendo}
\begin{figure}[ht]
    \centering
    \includegraphics[width=0.9\columnwidth]{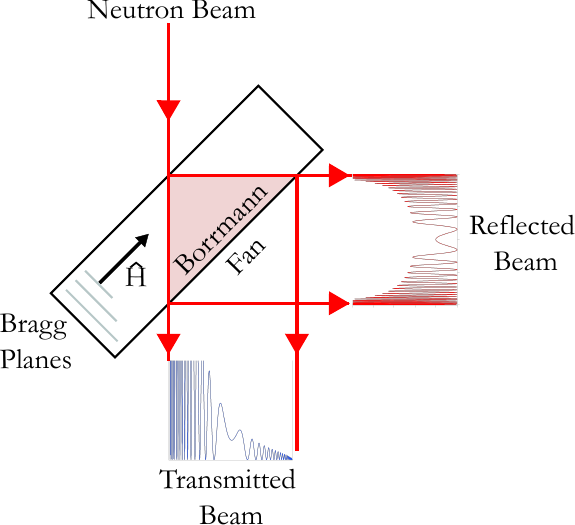}
    \caption{Diagram showing a crystal in the Laue geometry with the lattice planes perpendicular to the reciprocal lattice vector, \textbf{H}, and the transmitted and reflected beam directions. }
    \label{fig:laue_geo}
\end{figure}

Beyond utilizing perfect-silicon crystals to construct a closed-loop, MZ-like interferometer\textemdash where the interaction of interest is occurring between the crystal blades\textemdash interference effects also occur within the perfect crystals themselves.  
Known as \pendo interference,  this phenomenon results from  the neutron undergoing multiple scattering as it traverses the crystal, a process described by the theory of dynamical diffraction \cite{Rauch}.
While the details will be omitted here, the interference arises from the superposition of the two solutions of the Schr\"{o}dinger equation, labeled $\alpha$ and $\beta$, inside the crystal and results in rapidly oscillating intensity profiles like those shown in Fig. \ref{fig:laue_geo}.
The period of oscillation is given by the ratio of the crystal thickness $t$ over the \pendo length $\Delta_H$.  The \pendo length is determined by the crystal lattice and for silicon is typically on the order of \SI{50}{\micro m}.
The oscillations shown in  Fig. \ref{fig:laue_geo} are usually too fine to resolve in neutron scattering. 
To map out an interference curve, either the incident neutron wavelength is varied (scaling $\Delta_H$) or the effective crystal thickness is changed (scaling $t$).
For the work in Heacock et al. \cite{science1}, interference was observed by varying the effective thickness of the crystal for the (111), (220), and (400) reflections.
Precision measurements of \pendo interference give rise to determinations of the crystal's structure factor and several interesting physics results.

\par
The neutron-nucleus interaction in the crystal structure factor is influenced by the thermal vibrations of each atom about its equilibrium position, effectively "smearing out" the spatial distribution of the atom. 
The thermal oscillations produce a reflection orientation dependence  parameterized by the Debye-Waller factor, denoted by W. 
The Debye-Waller factor modifies the effective scattering length in the following way
\begin{equation} \label{eqn:sf}
    b_c \rightarrow b_ce^{-W}
\end{equation}
where the Debye-Waller factor is approximated as
\begin{equation} \label{eqn:DW}
    W=\frac{B}{16\pi^2}|\mathbf{\hat{H}}|^2
\end{equation}
where $\mathbf{H}= 2\pi/d$ is the reciprocal lattice vector, $d$ is the spacing between lattice planes, and $B$ is the Debye-Waller parameter. 
The room temperature Debye-Waller $B$ parameter for silicon has been determined using a least-squares fit to a Born-von K\'arm\'an (BvK) model \cite{Flensburg1999} as $B = $ \SI{0.4725+-0.0017}{\square\angstrom} for $T=$ \SI{295.5}{K}, whereas Heacock \cite{science1} measured a value from \pendo measurements of  $B = $ \SI{0.4761+-0.0017}{\square\angstrom} at $T=$ \SI{295.5}{K}. 
\par
Temperature gradients and absolute temperature are sources of uncertainty in \pendo measurements. Temperature gradients cause different parts of the crystal to expand and contract unevenly which cause strain gradients. 
These strain gradients cause a phase shift in the \pendo interferograms. 
The absolute temperature of the crystal changes the Debye-Waller factor (Eqn.\ \ref{eqn:sf}) which has an effect on the structure factor measurements using \pendo interference. This temperature dependence is closely tied to crystal lattice dynamics \cite{Lovesey1984}.
\par
The \pendo measurements by Heacock \cite{science1} included systematic uncertainties due to the temperature gradients in the crystal and the absolute sample temperature. 
Temperature gradients induced phase shifts of \SI{0.2}{\degree}, \SI{0.4}{\degree}, and \SI{3.7}{\degree} while the absolute temperature induced phase shifts of \SI{-1.0}{\degree}, \SI{-1.0}{\degree}, and \SI{3.2}{\degree} for the (111), (220), and (400) reflections, respectively. 
Each of these phase shifts had relative uncertainties at the $10^{-5}$ - $10^{-6}$ level contributing to the total (statistical and systematic) relative uncertainty of $10^{-5}$ for each reflection.
Performing the \pendo measurements in a more temperature isolated environment could eliminate these systematic uncertainties and improve those measurements by \SI{5.42}{\%}, \SI{7.05}{\%}, and \SI{42.7}{\%} for the (111), (220), and (400) reflections, respectively.


\section{Facility}\label{sec:facility}

The \NCNR{} houses a \SI{20}{\MW} research reactor that produces cold and thermal neutrons for research \cite{NCNRAnnualReport} including fundamental neutron physics, polymers, biological materials, materials science, and superconductivity. 
Neutrons are delivered to the approximately 30 neutron scattering instruments via highly reflective neutron guides (NG). 
The \NCNR{} is home to a state-of-the-art facility for neutron interferometry experiments with two independent monochromatic beamlines \cite{Rauch}. 
The oldest, the \NIOFh, was built around 1994 and consists of a series of nested enclosures.
The \NIOFh{} was designed specifically for perfect-crystal neutron interferometry by eliminating known sources of systematics. 
The outer most enclosure is constructed from concrete blocks that allows for some environmental isolation from the surrounding neutron instruments. 
The inner, \qty{40000}{kg} enclosure allows for thermal and acoustic isolation and, most importantly, has an active vibration isolation system. 
Experiments are conducted inside a third volume inside the inner enclosure. 
It consists of a cadmium lined aluminum box that reduces neutron backgrounds and maintains a constant internal temperature to within \SI{+-5}{mK} \cite{TempCont, Shahi_2016_NIM} using PID feedback control. 
\par
The \NIOFa{} is an additional NI facility located directly upstream and on the same cold neutron guide, NG-7, as \NIOFh. 
It benefits from a higher flux (\SI{3d6}{\per\centi\square\metre\per\s}), improved neutron polarimetry capabilities, and greater ease-of-access, but lacks many of the environmental controls and long term phase stability provided by the \NIOFh \cite{Shahi2016}. 
The \NIOFa{} was built to offer a more accessible experimental space for perfect-crystal neutron interferometry experiments \cite{Shahi_2016_NIM}. 
Vibrational stability can be improved through the use of a decoherence free subspace (DFS) neutron interferometer \cite{DFNI,DFS_NI}, which provides the interferometer some vibrational insensitivity. 
Thermal isolation can be achieved by placing the interferometer crystal inside a vacuum vessel \cite{Saggu2016} which is discussed below (See Sec.~\ref{sec:Olympus}). 
In addition, the open floor space of \NIOFa\ allows for the use of larger support apparatus such as a sample cryostat (discussed in Sec.~\ref{sec:cryo}).
Despite the additional noise, the \NIOFa{} has been used to investigate subsurface damage in thick-crystal neutron interferometers \cite{Heacock2019} and demonstrate far-field interference of a  bi-chromatic neutron beam \cite{Pushin2017}.
Most recently, \pendo interferometry has been used at \NIOFa{} to measure the neutron charge radius, silicon lattice dynamics, and constraints on a beyond the standard model 5\textsuperscript{th} force \cite{science1}.

\section{Vacuum Chamber: \emph{Olympus}}\label{sec:Olympus}

\subsection{Introduction}\label{sec:Olympus:intro}

Perfect-crystal neutron interferometry experiments are conducted at or near standard temperature and pressure (STP). 
Air scattering, temperature effects, and humidity changes are sources of correction and  uncertainty in these measurements. 
The phase shift in radians due to the nitrogen and water molecules in the air (Eqn.~\ref{eqn:Phase}) is approximately given by 
\begin{eqnarray}
    \Delta\phi_\textrm{air} = \big(-0.4 \times 10^{12} \; \text{m}^{-2}\big)\cdot D\cdot\lambda
\end{eqnarray}
with humidity changes making at most a \SI{3}{\%} correction in $\Delta\phi_\textrm{air}$ \cite{Rauch}.
For a recent example, air scattering caused a \SI{105.3+-0.8}{\degree} correction and approximately a quarter of the systematic uncertainty in the normalization of the \pendo results reported by Heacock \cite{science1}.
\par
Even small differences in temperature between the samples and crystal blades can cause noticeable phase shifts; a process so-called the shadow phase and discussed in Haun et al \cite{Haun2020}.

The use of a small vacuum chamber with two heating elements (one external and one internal) has been shown to reduce these effects \cite{initial_vacuum}. 
In Saggu \cite{initial_vacuum}, the temperature at the base of the interferometer inside the chamber was stable between \SI{24.400}{\degreeCelsius} and \SI{24.401}{\degreeCelsius}. 
The contrast was $\mathcal{C} = $ \SI{49.2+-2.1}{\%}  in vacuum (without pumping) compared to  $\mathcal{C} = $ \SI{46.0+-2.0}{\%} while the chamber was at STP. 
Introducing a small vacuum vessel most impacted phase stability.
The phase was stable to \SI{0.003+-0.03}{\degree\,per\,hour} over a 3-day period where no appreciable phase drift was observed. 
Under similar conditions, a peak-to-peak phase drift of \SI{140}{\degree} was observed during a 3-day period without use of the vacuum chamber \cite{initial_vacuum}.
\par
This paper describes a vacuum chamber vessel named \emph{Olympus} that has been added to the \NIOFa{} to help stabilize the environment in which these experiments are conducted, eliminate entirely air scattering effects, minimize changes to the Debye-Waller factor, and allow for the use of cryogenic samples. 
The chamber has been vacuum tested and holds a constant pressure of \SI{0.1}{mPa} (\SI{0.8d-6}{torr}). 
The lowest pressure achieved during the initial tests was \SI{0.039}{mPa} (\SI{2.9d-7}{torr}). 
\par
In early 2021, the \NCNR{} reactor had an incident involving an unlatched fuel element \cite{Reactor} resulting in a pause of operations.  
Since then the \NCNR\ reactor has not provided neutrons for scientific purposes. 
Meanwhile, the facility has undergone repairs, updated training, procedural reviews, and instrument upgrades.
In the near future, once the \NCNR{} resumes operations, the vacuum chamber will be installed at \NIOFa{} for perfect crystal neutron interferometry.

\subsection{Overall Design}\label{sec:Olympus:Flange}

The goal of the overall design of the chamber was to offer a high degree of flexibility in composition and placement of the internal components of the chamber in order to allow for its use in a variety of experiments. 
The top flange has multiple smaller flanges to provide access to the internal components of the chamber. 
There are four interchangeable side flanges with different layouts. 
The flange placement and interchangeability allows for many different configurations.

The chamber accommodates the use of a cryostat for cold samples.  The design includes two thin aluminum neutron windows to transmit a neutron beam.
A custom aluminum breadboard  suspended from  the top flange  allows for  access to the experimental setup inside the chamber. 
An isolation bellows coupler minimizes vibration to the chamber from the vacuum pumps and external vibrations. 
Vibrations introduce a phase shift between the two paths of ($a\cdot \mathbf{H})\tau_0^2$ where $\tau_0$ is the transit time of the neutron in between two blades and $a$ is the acceleration \cite{Bauspiess_1978_Nucl,Arif_1994_Vibr}.  
 
\begin{figure}[ht]
    \centering
    \includegraphics[width=0.9\columnwidth]{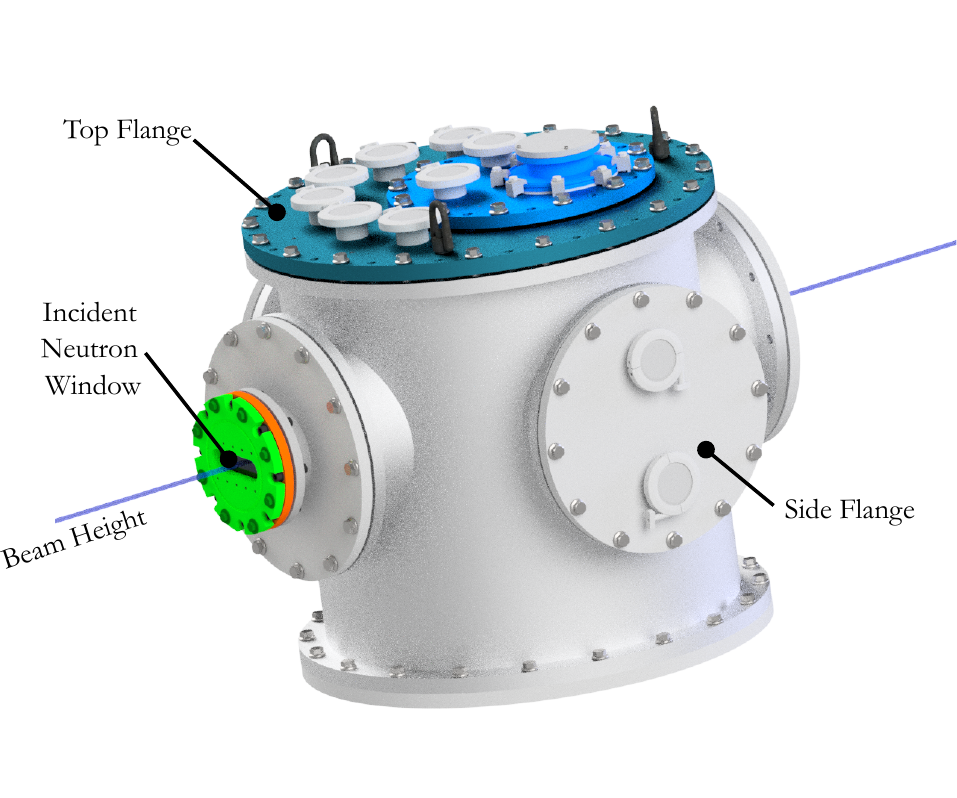}
    \caption{A rendered side view of the vacuum chamber \emph{Olympus}.}
    \label{fig:full_chamber}
\end{figure}

The body of the chamber is a cylinder with an internal height of \SI{673}{mm}, an internal diameter of \SI{648}{mm}, and wall thickness of \SI{6.35}{mm}. 
The distance from the base of the chamber to the center axis of the side flanges is \SI{424}{mm}. 
The chamber is supported by two guide rails mounted on a \SI{1.22}{m} $\times$ \SI{1.52}{m} optical table that has \sfrac{1}{4}-20 UNC [U.S. Customary Units] tapped holes located every \SI{25.4}{mm}.  The rails allow the chamber to move perpendicular to the neutron beam. 
While mounted on these rails the neutron beam is nominally at a height of \SI{511}{mm} from the surface of the optical table. 
The total internal volume of the chamber is \SI{236}{L}. 
Pressure inside the chamber is monitored by two pressure sensors, an ion gauge and wide range gauge. 
See Fig. \ref{fig:full_chamber} for a side view of the full chamber design.

\par 
\emph{Olympus} was designed with the following criteria:
\begin{itemize}
\item The use of standard vacuum components (ISO-K/F) wherever possible.
\item The chamber body and internal components are made from aluminum whenever possible to minimize magnetic interactions and material activation from the neutron beam.

\item Incorporate the use of an ion pump directly coupled to the chamber to avoid the vibrations caused by a turbo pump. To extend the life of the ion pump, a roughing pump combined with a turbo pump is used to achieve the lowest possible starting pressure for the ion pump, \SI{0.13}{mPa} (\SI{1d-6}{torr}). 
\item The use of thin windows to allow the neutron beam to pass through with minimal material scattering while maintaining the vacuum. 
\item Ability to power and manipulate multiple precision stages inside the chamber to accommodate different experiments.
\item Temperature control to maintain a constant temperature inside the chamber \SI{+-0.005}{\degreeCelsius} while also minimizing the internal temperature gradient.
\item Multiple viewports to visually monitor the experimental apparatus inside the chamber.
\item A platform attached to the top of the chamber that will hold all the experimental components, lift out of the chamber when the top is lifted, and facilitate experiment flexibility and access to experimental components.
\end{itemize}

\subsubsection{Top Flange}\label{sec:Olympus:TopFlange}

\begin{figure}[ht]
    \centering
    \includegraphics[width=0.9\columnwidth]{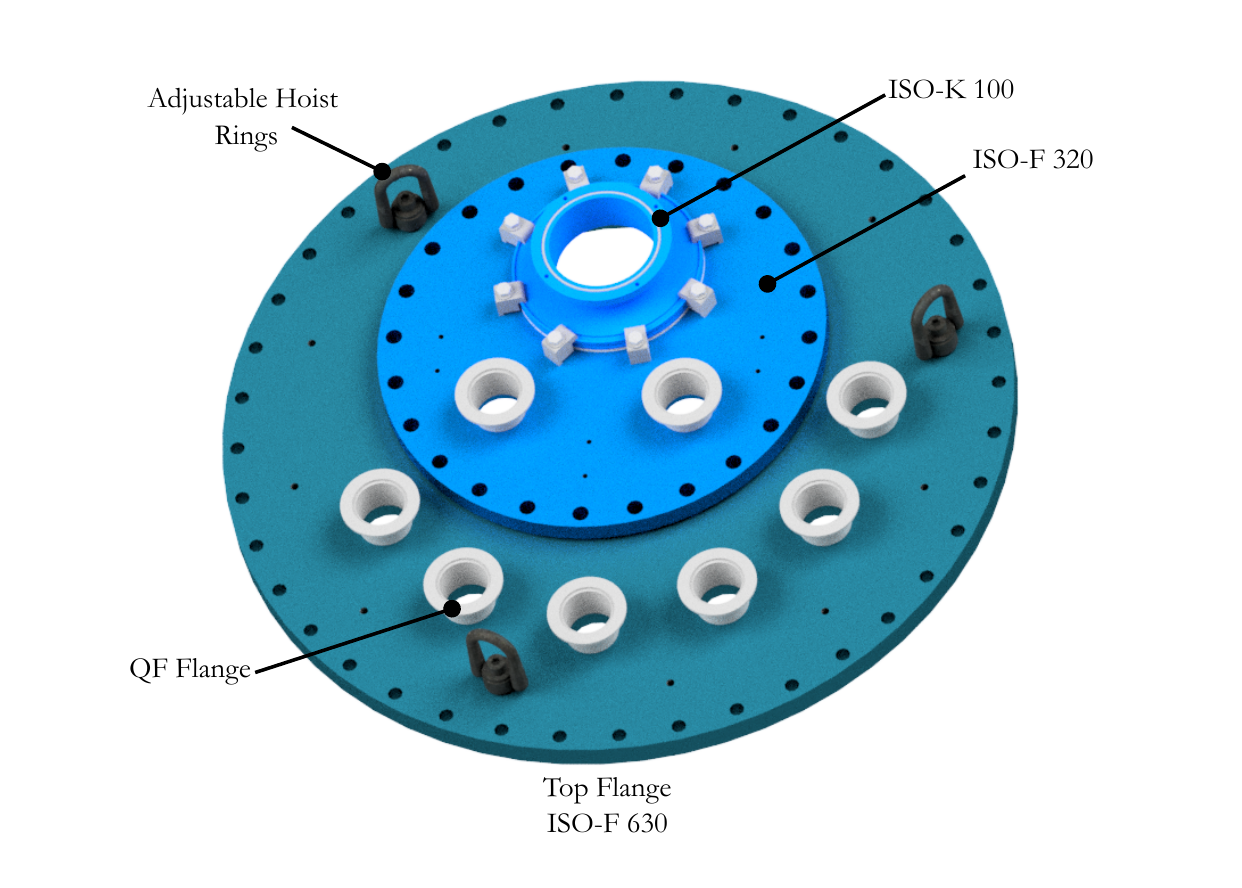}
    \caption{The top flange has several welded connections for feedthroughs and other equipment.  An ISO-K\,100 flange allows the connection of a cryostat. }
    \label{fig:top_flange}
\end{figure}

The top flange, or ``lid", (See Fig. \ref{fig:top_flange}) of the chamber is an ISO-F\,630 flange modified to have 40 bolt holes (only 20 bolts are needed to secure the flange) to allow the lid to be rotated in \SI{7.5}{\degree} increments. 
The lid has one off-axis ISO-F\,320 flange which was modified to have twice the number of standard bolt holes, similar to the ISO-F\,630, for finer rotational adjustments. 
The lid also has six welded QF-50 flanges arranged in a semicircle around the ISO-F\,320 flange.
The QF-50 flange placement was designed to offer the most flexibility for feedthroughs into the system for electrical, lighting, windows, gauge placement, and to allow a high degree of flexibility in experimental setup. 
There are 12 blind, \nicefrac{1}{4}-20 tapped holes arranged on a \SI{620}{mm} diameter circle for hoist rings to be attached.
This allows for the lid to be lifted by crane for access to the internal components without interfering with any experimental equipment attached to the QF-50 flanges.  
\par
The ISO-F\,320 flange was designed to be interchangeable with different flange designs in order to increase variability without the need to design a new ISO-F\,630 top flange for every experiment. In one such configuration, the ISO-F\,320 has an ISO-K\,100 flange with two additional QF-50 flanges (see Fig.~\ref{fig:top_flange}).
The ISO-K 100 flange allows for a cryostat to be attached to the chamber for cooling NI samples while the QF-40 flanges are for any necessary feedthroughs and viewports. 

The ISO-F\,320 flange is off center to allow for the highest number of unique sample positions at the end of the cryostat.
By rotating the top flange of the chamber and the smaller ISO-F\,320 flange separately the sample position of the cryostat has over 800 unique positions inside the chamber. 
\subsubsection{Side Flanges}\label{sec:Olympus:SideFlange}

The four side flanges are all ISO-F\,320, \SI{90}{\degree} apart and nominally centered on the neutron beam height at \NIOFa.
These flanges can be customized and interchanged as they are made from standard ISO flanges. 
Currently, two side flanges are used for neutron windows (discussed next) while the ones perpendicular to the beam can have either QF-50 flanges (for feedthrough access) or an axially centered ISO-K\,160 flange (for access and pumping attachments).  
These initial configurations were designed for the first experiments to be conducted with the chamber.

\subsubsection{Window Flanges}\label{sec:Olympus:Window}

The incident neutron window has been designed to allow the neutron beam to pass into the chamber through the least amount of material possible in order to minimize neutron scattering and absorption. 
We utilized the cost effective technique of repurposing  aluminum beverage cans as a neutron window material \cite{aCORN}.
Aluminum cans are manufactured by stretching aluminum disks to a thickness of only \SI{0.11}{mm}.  
\begin{figure}[ht]
    \centering
    \includegraphics[width=0.9\columnwidth]{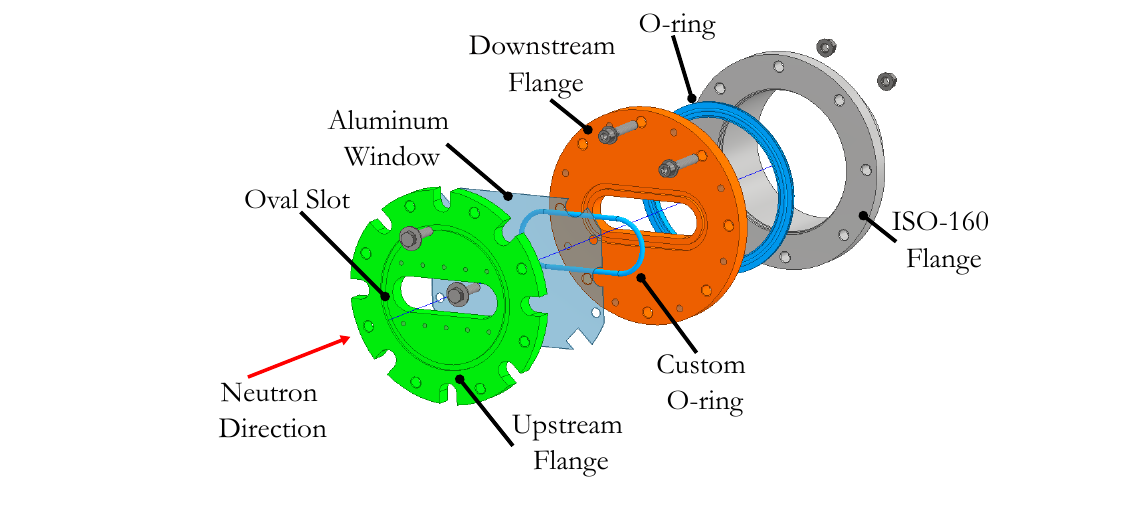}
    \caption{Assembly of the incident neutron window frame. An aluminum window made from a commercial beverage can is sandwiched between 2 modified ISO-F 100 flanges and then attached to an ISO-160 flange mounted on a side flange (ISO-F 320) of the vacuum chamber.  Most fasteners have been removed for clarity.}
    \label{fig:window}
\end{figure}
As shown in Fig.\ \ref{fig:window}, a rolled beverage can is pressed between two ISO-F\,160 flanges. 
A center oval entrance slot, \SI{38}{mm} tall and \SI{124}{mm} wide, was machined through both ISO-F\,160 flanges.
The exposed part of the beverage can window is larger than the typical neutron interferometer experiment which uses a neutron beam $\leq$ \SI{12.7}{mm} in diameter.  
The horizontal length of the oval is large since the vacuum chamber may move perpendicular to the beam depending on the specific experiment setup. 
Ten tapped holes (five above and five below) surround the oval slot to allow for the mounting of collimation slits.
A custom o-ring groove creates the vacuum seal against the window material. 
\par
Since the O-beam is displaced from the incident beam by an offset that may differ based on experimental parameters, a different approach was used for the neutron exit window on the most downstream side flange.
An ISO-F\,160 blank was milled so that there is a centered \SI{35}{mm} tall by \SI{254}{mm} wide slot.  
This slot was by created by removing \SI{18}{mm} from the internal surface of the flange creating a window approximately \SI{2}{mm} thick. 
\par
For the outgoing H-beam, exiting the interferometer at a $2\theta_B$ angle close to \SI{90}{\degree}, the beam exits through an ISO-F 160 blank. 
This blank is attached to \emph{Olympus} via a Tee flange with two ISO-F 160 connections and one ISO-F 100 connection. The opposite side of Tee flange is connected to an ISO-F 320 side flange. The narrower ISO-F 100 part of the Tee flange is connected to the vibration isolation components described in Sec. 4.4.  This setup is shown in  Fig.\ref{fig:vibration_isolation}). One side of tee flange is connected to an ISO-F\,320 side flange and the other side has the exit window blank. The ISO-F\,100 side of the tee flange is connected to the vibration isolation components described in Sec. \ref{sec:Olympus:tvibration}.
The H-beam exits from the ISO-F\,320 end of the tee flange. 
The majority of the incident beam that does not satisfy the Bragg condition  will be absorbed on a beam block after the interferometer crystal but before leaving the chamber.

\subsection{Internal Components}\label{sec:Olympus:internal}

\begin{figure}[ht]
    \centering
    \includegraphics[width=0.9\columnwidth]{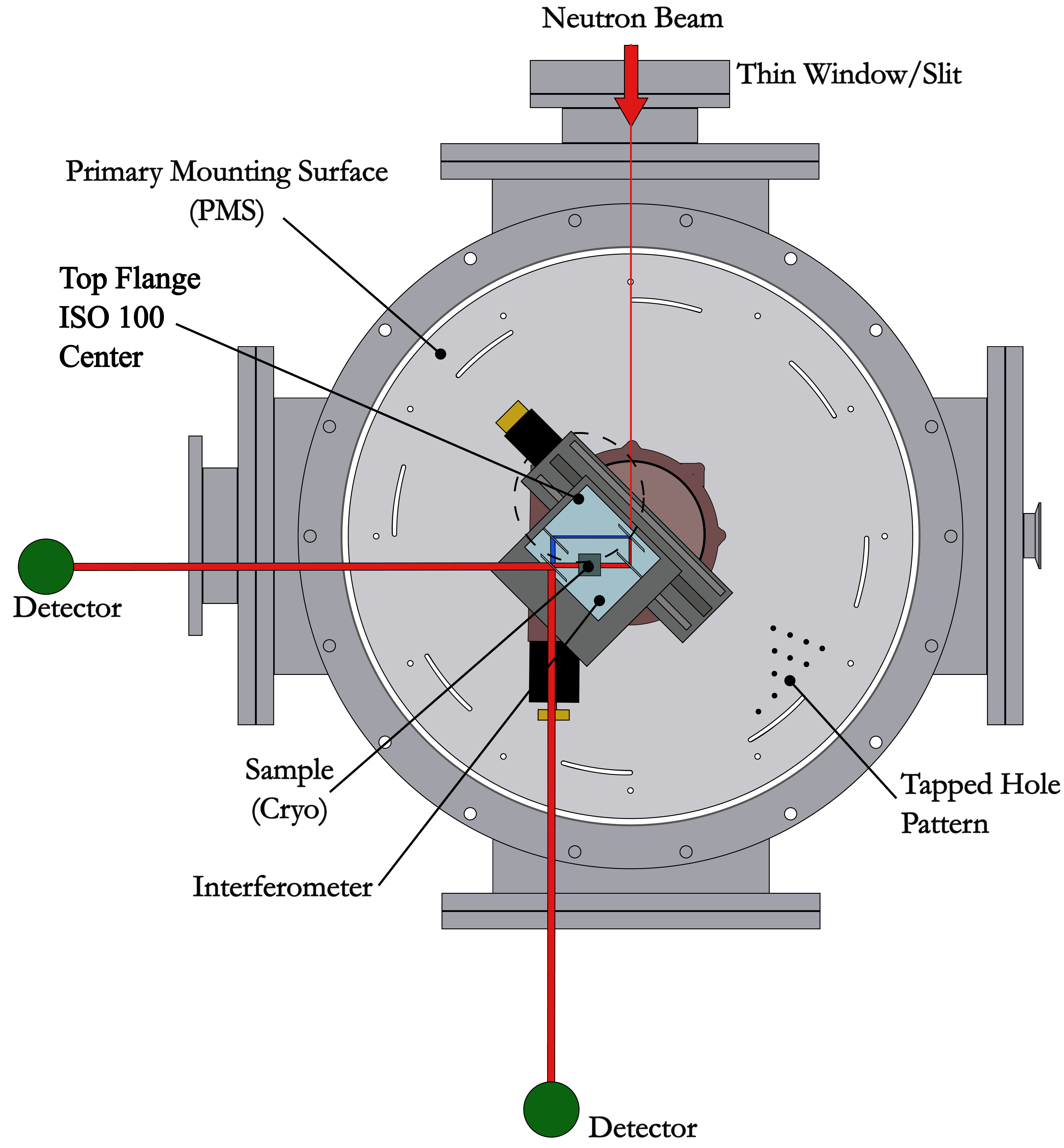}
    \caption{Top down schematic of the inside of the vacuum chamber showing the placement of the translation stages, interferometer, incident neutron beam, and neutron beam path inside and after the interferometer. The tapped hole pattern repeats throughout the PMS}
    \label{fig:stack_schematic}
\end{figure}

The internal side of the top flange has 24, \sfrac{1}{4}-20   UNC [U.S. Customary Units] blind-tapped holes that allow up to twelve  aluminum support rods,
\SI{648}{mm} long by \SI{12.7}{mm} in diameter,  to be attached.
These support rods suspend a \SI{266.8}{mm} diameter circular aluminum plate \SI{18}{mm} above \emph{Olympus}' inner bottom surface.
The plate has a \nicefrac{1}{4}-20 tapped through-hole pattern spaced every \SI{25.4}{mm}$\times$ \SI{25.4}{mm} and serves as the primary mounting surface (PMS) for  experimental components. 
Eight \SI{81}{mm} long countersunk slots equally spaced on the circular plate allow the aluminum plate to be bolted to the internal side of the bottom surface of the chamber for stability, if needed. 
The purpose of the PMS is to be flexible for multiple experimental setups, minimize outgassing, and to allow the experimental setup to be vertically lifted out of the chamber while not disturbing the integrity of the configuration required for the experiment. 
\begin{figure}[h]
\centering
\includegraphics[width=0.9\columnwidth]{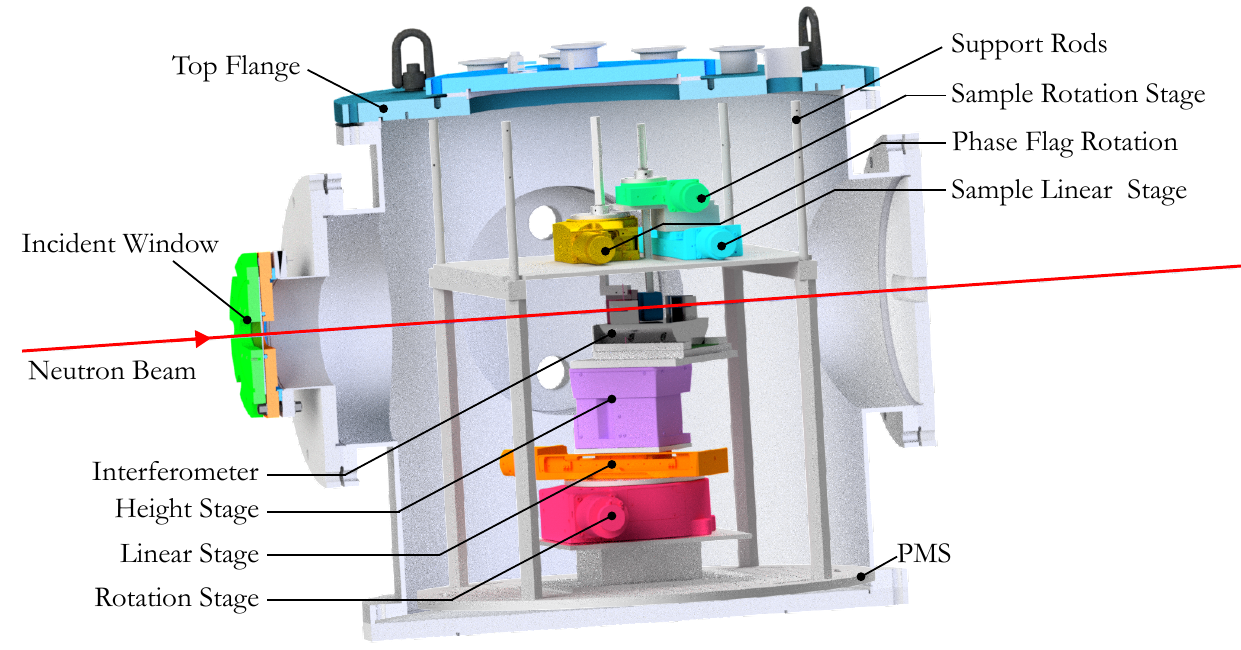}
\caption{An example configuration of \emph{Olympus} when not using the cryostat. Support rods suspend a Primary Mounting Surface (PMS) from the top flange.  A stack of stages allows for the interferometer to be aligned to the incident neutron beam. A sample and phase flag are mounted above the interferometer.}
\label{fig:full_basket}
\end{figure}
Fig.~\ref{fig:full_basket} shows the PMS, support rods, and top flange of the chamber. 
Three vacuum compatible stages, collectively referred to as the ``stack", were  installed into the chamber to support and align the interferometer crystal. 
The stack consists of a Physik Instrumente (PI) L-310 Z (height) stage, a PI PLS-85 linear stage (nominally along $\hat{x}$, parallel to the crystal blade surface), and a PI PRS-200 rotation stage \cite{NISTDisclaimer}.
The stages have a travel range of \SI{26}{mm}, \SI{155}{mm}, and \SI{360}{\degree}, respectively. 
Likewise, the resolution of the stages are \SI{0.3}{\micro\meter}, \SI{50}{nm}, and \SI{7.627118d-5}{degrees}. 
Fig.~\ref{fig:stack_schematic} shows a top down schematic of the stack inside the chamber.
\par
Non-cryogenic samples and the phase flag are supported by two \SI{25.4}{mm} square aluminum bars (\SI{432}{mm} long) attached to the aluminum support rods that hold the PMS. 
These bars suspend an aluminum plate above the stack and interferometer. 
The plate fastens to the square aluminum bars and can be modified for different experiments.
\par
The plate holds translation and rotation stages for the phase flag and sample. 
The phase flag is rotated by a PI PRS-110 precision rotation stage with a resolution of \SI{0.0002}{\degree}. 
A sample is rotated by a PI DT-80 compact rotation stage with a resolution of \SI{0.01}{\degree}  and can be translated in and out of the interferometer using a PI PLS-85 precision linear stage.

\subsection{Vibration Considerations}\label{sec:Olympus:tvibration}

\begin{figure}[ht]
    \centering
    \includegraphics[width=0.9\columnwidth]{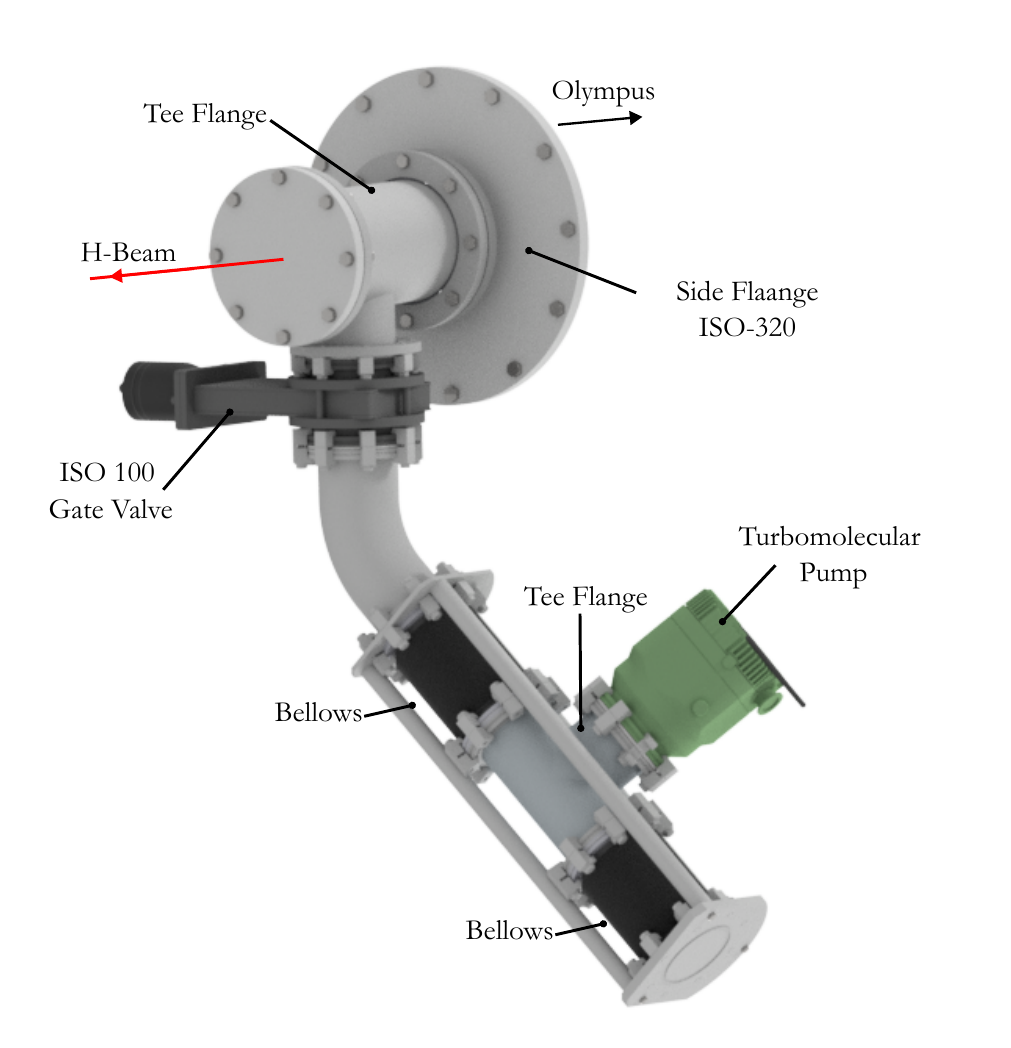}
    \caption{The isolation system that lessens the turbo pump vibrations from reaching the vacuum chamber. The turbo pump is supported independently (not shown). }
    \label{fig:vibration_isolation}
\end{figure}

 Perfect-crystal neutron interferometry is very sensitive to vibrational noise. 
 In order to achieve the lowest starting pressure to activate an ion pump, the system uses a roughing pump followed by a turbomolecular (turbo) pump before switching to an ion pump to limit vibrations.
The optical table sits on rubber pads that provide some vibration damping from the surrounding building but not at the level of suppression required for an operating turbo pump.
To isolate vibrations caused by the turbo pump to the rest of the system, the turbo pump is fastened to a free standing support that is separate from the optical table below the vacuum chamber. 
The ISO-K\,100 flange on the pump is connected to the middle arm of a tee flange whose two sides are connected to two flexible bellows hoses (see Fig.\ \ref{fig:vibration_isolation}) that dampen the vibrations from the pump.
 The supporting structure is then connected to the chamber by a tee flange with two ISO-F\,160 connections and one ISO-F\,100 connection. 
 The tee flange was selected so that a diffracted H-beam can pass through this connection while the turbo pump is attached and operating. 
 The ISO\,100 end of the tee flange is connected to a gate valve that can isolate the pump from  \emph{Olympus}. 

\section{Cryostat}\label{sec:cryo}

\subsection{Description}\label{sec:cryo:descipt}

\begin{figure}[h]
\centering
\includegraphics[width=0.9\columnwidth]{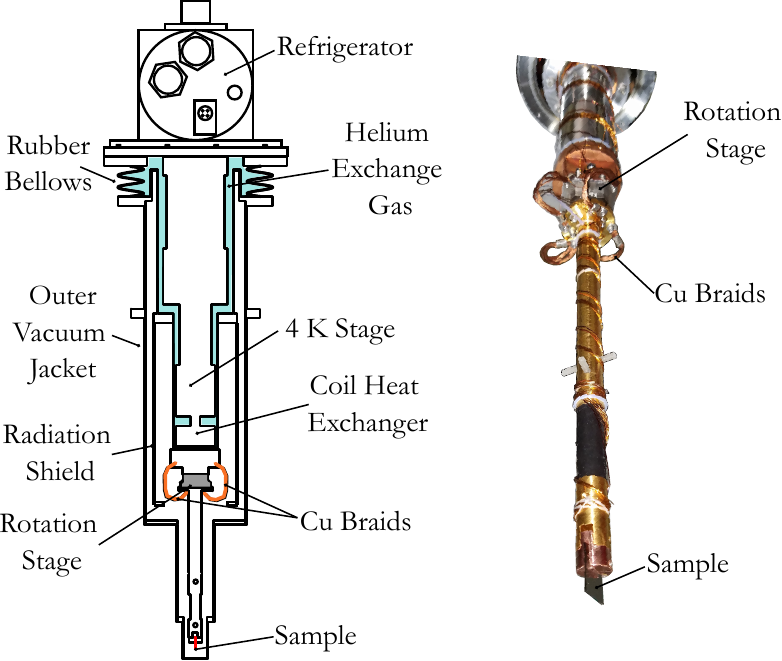}
\caption{(left) The cryostat for running cryogenic samples inside a neutron interferometer. Note that the rubber bellows prevents vibrations from the refrigerator from influencing the sample and interferometer crystal. Copper (Cu) braids that attach the cold head to the sample holder, bypassing the rotation stage, allow for more efficient transfer of heat loads.  (Right) The cold finger and sample with the outer vacuum jacket removed. }
\label{fig:cryostat}
\end{figure}

The cryostat used to cool samples is a model SHI-4XGS-5 from Janis Research Company which operates from \SI{4.5}{K} to \SI{325}{K}. 
The cryostat utilizes a Gifford-McMahon-type refrigerator which has a cooling power of \SI{0.5}{\W} at \SI{4.2}{\K}. 
The cryostat is specially designed to vibrationally isolate the cryogenic sample from the refrigerator and cold head. 
This is critical in order to preserve coherence in the interferometer as the reflected neutron path receives a phase shift from the displacement of the lattice planes.
The vibration isolation is accomplished through the use of an exchange gas chamber. 
The cold head is surrounded by this exchange gas chamber, as shown in Fig. \ref{fig:cryostat}, with the sample holder mounted to the underside of the chamber. 
Helium exchange gas at a pressure of around \SI{101}{KPa} (1 atm) couples the cold head to the sample holder. 
A radiation shield and an outer vacuum jacket surround the exchange gas chamber. 
A rubber bellows located near the top of the cryostat isolates the upper portion of the cold head and allows the cold head to be supported  independently. 
At the \NIOFa, the refrigerator is supported by an aluminum frame  built around, but supported separately from, the optical table on which the interferometer and \emph{Olympus} is mounted. 
This frame allows limited motion of the cryostat in the X, Y, and Z directions. 
\par
During transport and positioning of the cryostat, the upper and lower sections are rigidly connected via temporary posts and brackets around the rubber bellows. 
To ensure that the lower portion of the cryostat remains stable during the course of the runs, a cylindrical clamp is installed around the outer vacuum jacket of the cryostat. 
This clamp is mounted to the optical table that supports the interferometer crystal. 
With the clamp in place, the temporary brackets and posts are removed and the upper portion of the cryostat is raised. 
This separates the two sections and establishes the vibrational isolation. 
\par
The sample to be investigated is mounted at the bottom of a long cold finger, which in turn is mounted on a cryogenic rotation stage with a resistive encoder (Attocube ANR 101/RES). 
The rotation stage is attached to the sample holder block. 
Because the rotation stage is not a good thermal conductor, a set of copper braids connects the sample holder block to the cold finger. 
\par
In order to fit in the relatively small gap between the interferometer blades, the lowest portion of the cryostat outer vacuum jacket is made as small as possible. It is constructed of \SI{1}{\mm} thick aluminum and has a square cross-section.
This minimizes the amount of neutron scattering and attenuation caused by the jacket. 
The sample cold finger contains a \SI{50}{\ohm} wire-wound heater that allows the sample temperature to be controlled over the entire range from \SI{4.2}{\K} up to \SI{325}{\K}. 
Two temperature sensors are mounted on the cryostat, one located at the sample holder block and one located on the cold finger, just above the sample location. 
\begin{figure}[h]
\centering
\includegraphics[width=0.9\columnwidth]{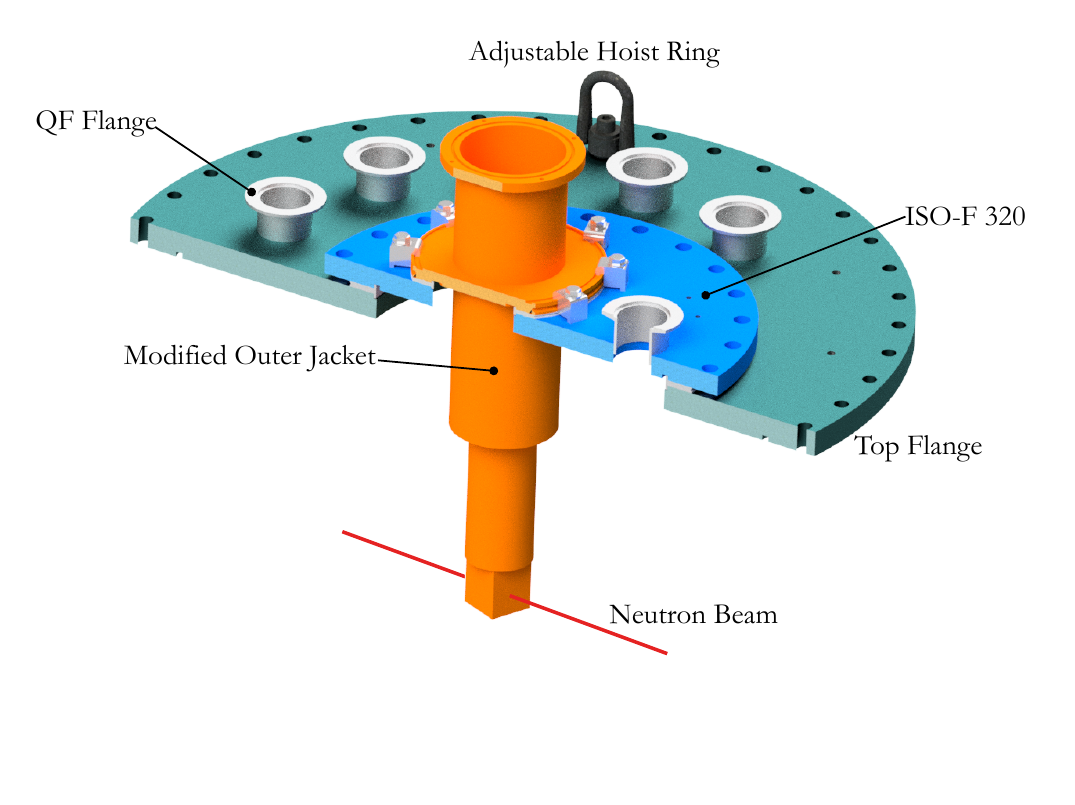}
\caption{A cross-section view showing a modified outer jacket that enables the cryostat to be coupled to \emph{Olympus}.}
\label{fig:CryoCoupled}
\end{figure}
\subsection{Cryostat Initial Tests}\label{sec:cryo:expt}
\begin{figure}[h]
\centering
\includegraphics[width=0.9\columnwidth]{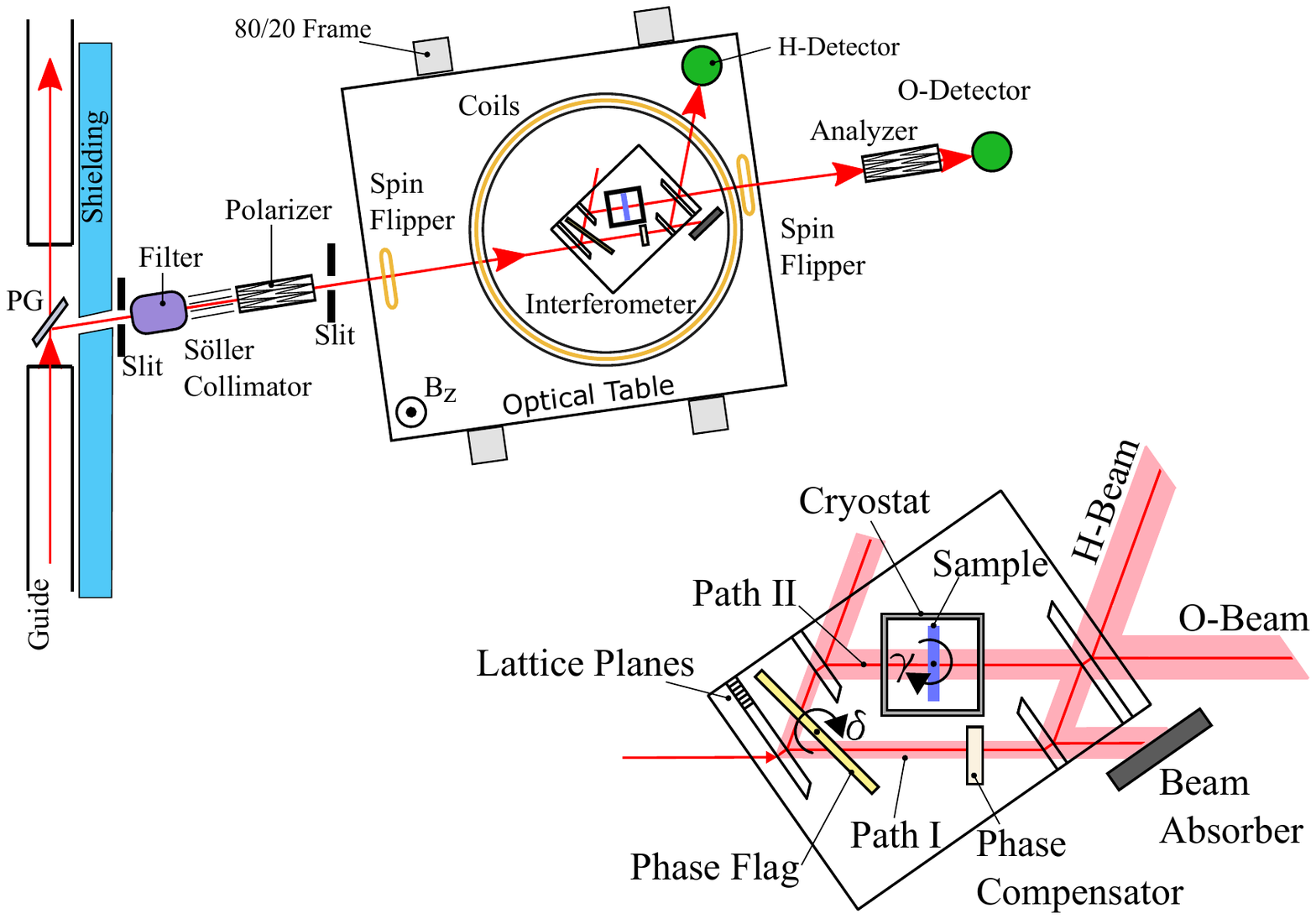}
\caption{Diagram of the interferometer facility at NIOF\textbf{a}.  Details are in text. }
\label{fig:sampleininterferometer}
\end{figure}

Initial testing of the cryostat was performed at the NIOF-a, without use of the vacuum setup.
Neutrons with a wavelength of \SI{0.44}{\nm} are delivered to the \NIOFa{} by a single reflection off of a pyrolytic graphite monochromator located in NG-7. 
Fig. \ref{fig:sampleininterferometer} shows a schematic of the facility.  
After reflecting off the monochromator, several slits and a s{\"o}ller collimator are used to limit the size and divergence of the neutron beam. 
A  beryllium filter is used to remove any higher energy neutrons ($\lambda < $ \SI{4.0}{\angstrom})  from reaching the apparatus. 
Neutrons can be polarized in the vertical direction  using a supermirror polarizer before entering the interferometer. 
The sign of the polarization state can be changed using a spin-flipper.   
Various permanent magnet assemblies along with Helmholtz coils provide the vertical magnetic field $B_z$ needed to preserve the neutron polarization direction as it traverses the apparatus.
The neutron interferometer itself rests upon an optical table surrounded by a cadmium-lined aluminum enclosure. 
Neutrons exiting the NI are detected using two $^3$He proportional counters.    
The neutron's polarization state, $P_n$, can be post-selected using supermirror analyzers. 
\par
For this work, a skew-symmetric interferometer (like shown in Fig.~\ref{fig:ni}) was utilized because it contains two long parallel paths and the largest gap for inserting the cryostat tip.
Unfortunately, the largest available skew-symmetric interferometers at NIST were machined to use the (220) lattice vector, as opposed to (111), which has a Bragg cutoff of $\lambda=$ \SI{3.8}{\angstrom}.  
Therefore, the beryllium filter at \NIOFa\ was removed and the experiment used the remaining $\lambda/n$ component of the incident beam.
This decreased the Bragg diffracted incident neutron intensity by \SI{75}{\%}.

\subsection{Sample Region}

A Helmholtz coil assembly was constructed around the interferometer and cryogenic sample in order to provide a magnetic field to the sample,  as seen in Fig. \ref{fig:sampleininterferometer}.
These coils are wound from 10 gauge copper wire, with 140 turns per coil.
The resistance of each coil at room temperature is \SI{0.5}{\ohm}.
With a current of \SI{15}{\A}, the magnetic field at the center generated by the coils is \SI{24}{\milli\tesla}. 
The coils are cooled to \SI{30}{\celsius} by flowing a water/glycol mixture through copper tubing wrapped around the coils. 

The original sample chosen to test the capabilities of the cryostat and interferometer was a \SI{16.54}{\mm} by \SI{14.46}{\mm} rectangle of \SI{1}{\um} thick coating of \sample{} sputtered onto a \SI{520}{\um} thick silicon wafer.
The expected phase shift (Eqn.~\ref{eqn:Phase}) of the materials in the neutron paths are then
\begin{align*}
\Delta\phi_{\mathrm{sam}} &= - 0.18 \text{ rad}  && \mathrm{Ni}_{60}\mathrm{Cu}_{40} \\[1ex]
\Delta\phi_{\mathrm{Si}}  &= -23.73 \text{ rad} && \text{Si wafer} \\[1ex]
\Delta\phi_{\mathrm{Al}}  &= -92.89 \text{ rad} && \begin{tabular}{@{}l@{}} Cryostat \\ jacket walls \end{tabular}
\end{align*}
at $T = $ \SI{300}{K} and $\lambda = $ \SI{2.2}{\angstrom}.

\par
This sample has a known Curie temperature transition around \SI{200}{\K}. 
The testing was performed in two separate configurations. 
The first involved placing the cryostat with sample into path II of a skew-symmetric interferometer. 
A diagram of this setup can be seen in Fig. \ref{fig:sampleininterferometer}. 
The second testing configuration utilized a Ramsey-type  setup rather than an interferometer.
Tests were performed and data collected at two different times, separated by approximately 8 months. 
This was due in part to the need to fix a vacuum leak in the cryostat. 
The sample was aligned using the cryogenic rotation stage and measuring the count variation as a function of the sample's effective thickness $D/\cos(\delta)$, where $\delta$ is an offset angle.
\begin{figure}[h]
\centering
\includegraphics[width=0.9\columnwidth]{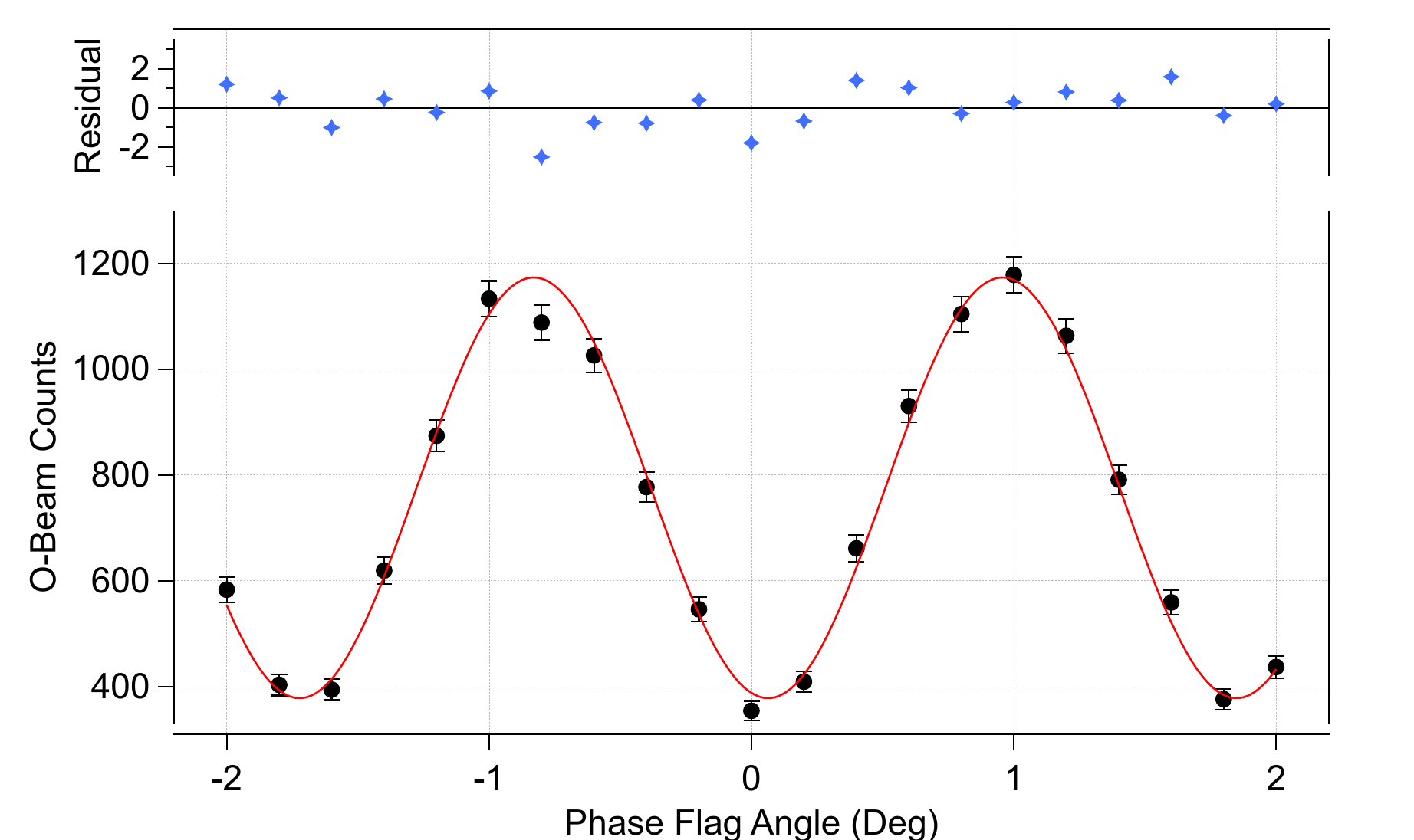}
\caption{Initial interferogram without the sample. Since the lower intensity $\lambda/2$  component of the neutron beam was used each point was taken for \SI{600}{s}. Residuals are defined as $(y_i-y_\mathrm{fit})/\sigma_i$. Uncertainties shown are purely statistical and are at the 68 \% confidence level. }
\label{fig:emptyininterferometer}
\end{figure}
\par
Initial tests of the sample and cryostat showed contrast (see Fig.~\ref{fig:emptyininterferometer}) of around $50\,\%$ for the sample in air; $28\,\%$ for the sample warm, inside the aluminum housing of the cryostat; and around $15\,\%$ with the sample cooled to \SI{15}{\K}. 
Utilizing the vibration isolation of the cryostat increased the cold contrast to around $24\,\%$ at \SI{14}{\K}. The measurements are summarized in Table \ref{tab:vibration}. Later runs saw contrast as high as $40\,\%$ at \SI{11.2}{\K}.
A phase compensator, consisting of a \SI{3}{mm} thick piece of aluminum, was  placed in the reflected beam path in order to compensate for the phase shift introduced by the cryostat's vacuum jacket walls.

Small pieces of neutron absorbing cadmium were used to block the direct beam and reduce background rates. 
The proximity of the direct beam and the O-beam meant that the observed contrast was fairly sensitive to the placement of this beam block. Both the placement of the phase compensator and Cd are shown in Fig. \ref{fig:sampleininterferometer}. 

\begin{table}
\caption{ Comparison of contrast under various configurations. Uncertainties are the weighted standard error of the mean.}

\begin{tabularx}{\columnwidth}{lcr}
\hline\noalign{\smallskip}
 &  Sample   &  Absolute   \\
 &   Temperature  &   Contrast  \\
 Condition &  (K) & (\%) \\
\hline\hline
None & 300 & 49.5 $\pm$~1.5 \\
\sample  & 300 & 51.4 $\pm$~2.6 \\
\sample{}  & \multirow{2}{*}{300}  & \multirow{2}{*}{ 29.2 $\pm$~3.0} \\
\hspace{0.5cm} + shielding & & \\
\sample{}   & \multirow{2}{*}{15} & \multirow{2}{*}{$17.0\pm 0.7$} \\
\hspace{0.5cm} + shielding & & \\
\sample & \multirow{3}{*}{14} &\multirow{3}{*}{$24.3\pm1.2$} \\
\hspace{0.5cm} + shielding & & \\
\hspace{0.5cm} + vibration iso. & & \\
& & \\
\hline
\end{tabularx}
\label{tab:vibration}
\end{table}

\par

\subsection{Interferometer}

\begin{figure}[h]
\centering
\includegraphics[width=0.9\columnwidth]{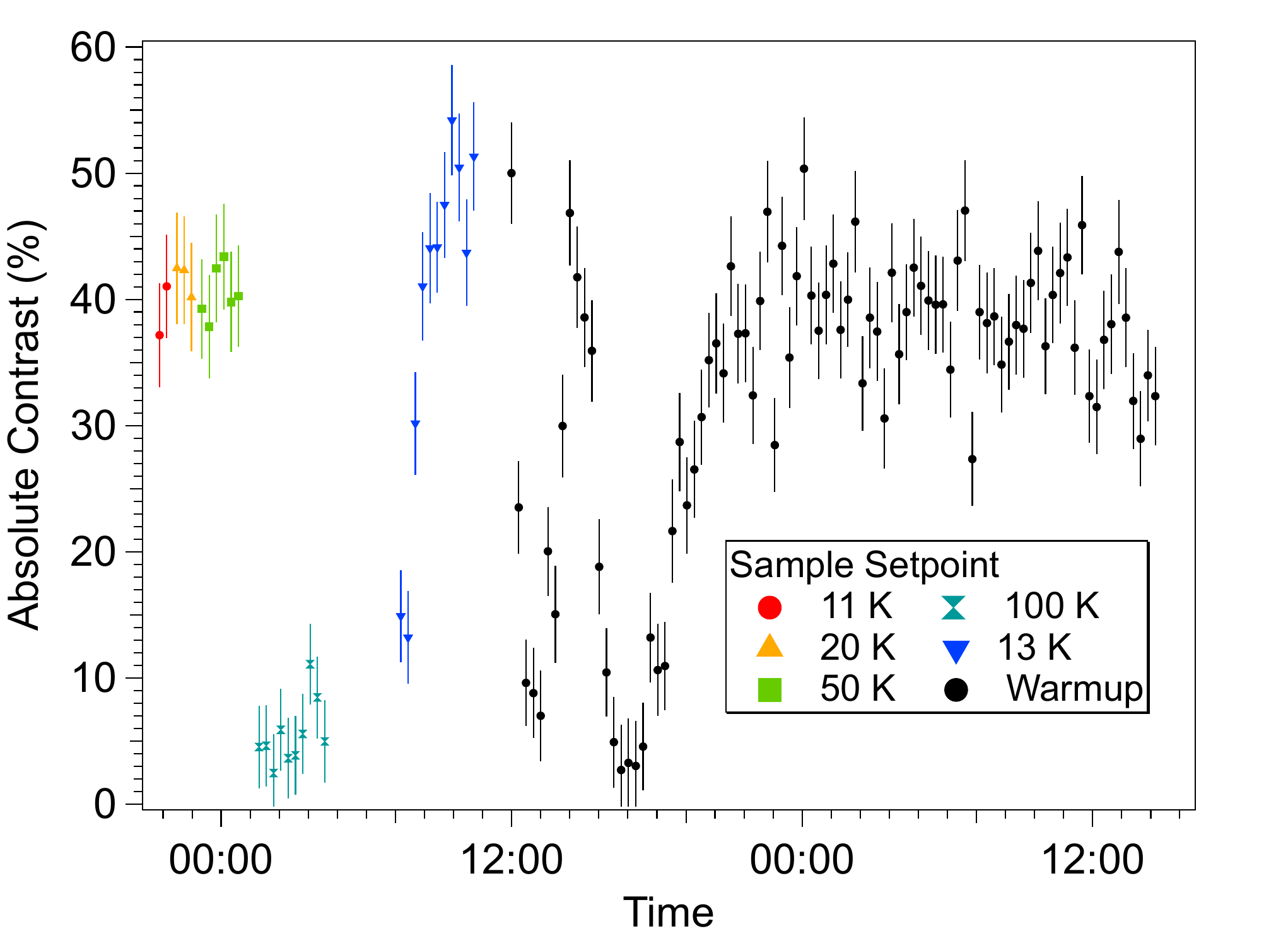}
\caption{Contrast measured while the sample was cooled inside the interferometer.  Various set points indicated by color/marker-type are shown.Uncertainties shown are purely statistical and are at the 68 \% confidence level. }
\label{fig:sampleAlignment1}
\end{figure}

\begin{figure}[h]
\centering
\includegraphics[width=0.9\columnwidth]{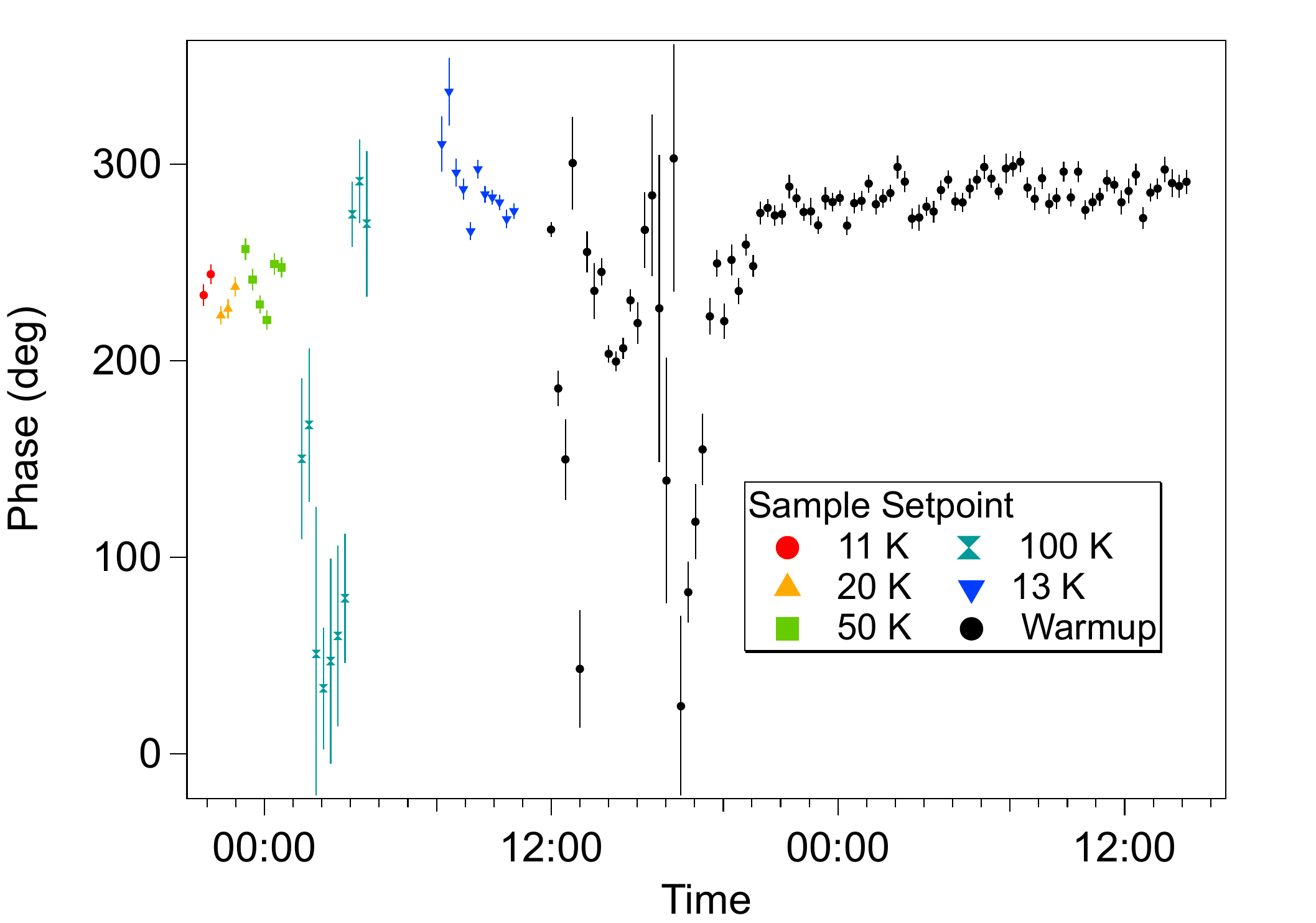}
\caption{Phase measured while the sample was cooled inside the interferometer for the same runs as in Fig.~\ref{fig:sampleAlignment1}. Various set points indicated by color/marker-type are shown.   Uncertainties shown are purely statistical and are at the 68 \% confidence level.}
\label{fig:sampleAlignment2}
\end{figure}

A typical experimental run involved establishing the contrast of the warm sample, proceeding to cool the sample to base temperature, and then increasing the temperature of the sample in steps back up to \SI{300}{\K}.  While early runs involved simply changing the temperature set point between the various steps, we found that slowly ramping the sample heater power (and thus the sample temperature) avoided spikes in the heater current and provided the consistent data. A ramp of \SI{2}{\K} per minute was used during later runs (see Figs. \ref{fig:sampleAlignment1} and \ref{fig:sampleAlignment2}). 
\par
We were able to operate and observe contrast across the full temperature range. 
However, we did not observe a repeatable change at the suspected transition temperature of the sample, nor at any other temperature due to that sample's limited thickness and the experiment setup.   

\par
\begin{figure}[ht]
\centering
\includegraphics[width=0.9\columnwidth]{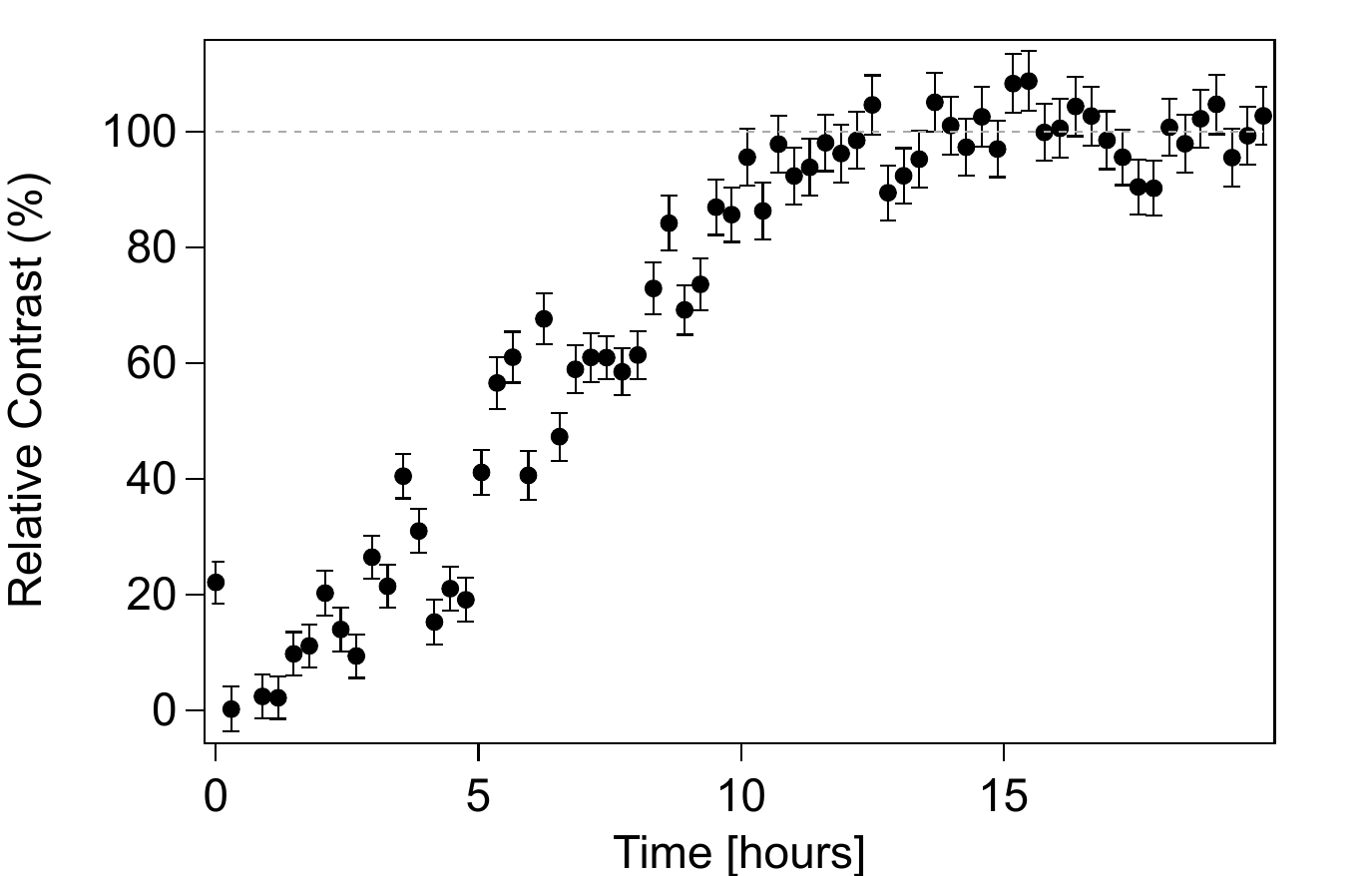}
\caption{Relative contrast of the interferometer as the sample warmed from \SI{11}{K} to room temperature.   Uncertainties shown are purely statistical and are at the 68 \% confidence level.  }
\label{fig:contrastRegained}
\end{figure}
\begin{figure}[h]
\centering
\includegraphics[width=0.9\columnwidth]{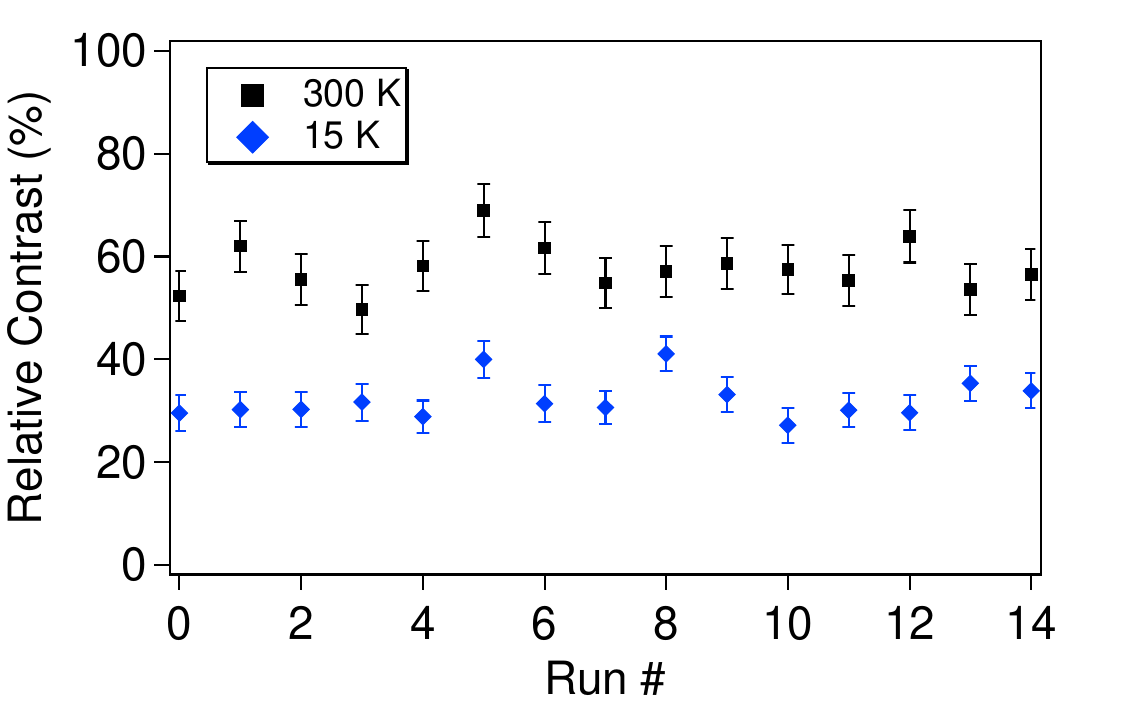}
\caption{Contrast with the \sample{} measured at 2 different temperatures relative to  empty interferometer of \SI{50}{\%} (See Fig. \ref{fig:emptyininterferometer}).  Uncertainties shown are purely statistical and are at the 68 \% confidence level. }
\label{fig:Relcontrast}
\end{figure}

During testing, we found that the temperature of the outer vacuum jacket of the cryostat varied during the course of operation. 
Due to radiative heat transfer between the cryostat walls and the interferometer, thermal gradients were induced across the crystal leading to a loss of contrast.
We demonstrated this by physically raising the cryostat  outer vacuum jacket just above the neutron beam.   The slightly cooler  outer vacuum jacket totally destroyed any contrast (see Fig.~\ref{fig:contrastRegained}) which was eventually recovered as the system naturally warmed up and thermalized to the environment.  
We found this effect could be minimized by wrapping a heater around the outside of the cryostat and using a PID loop to maintain the outside of the cryostat at a constant temperature (typically chosen to be \SI{22.5}{\celsius}). 
This preserved contrast in the interferometer.  
The relative contrast measurements with stabilized temperature are shown in Fig.\ \ref{fig:Relcontrast}.
 It is suspected that installing a radiation shield would have also solved this loss-of-contrast problem.

\section{Summary}\label{sec:summary}
Neutron interferometry is an extremely sensitive tool that can be used to measure nuclear and magnetic interactions in matter. 

Adaptive environmental isolation is needed for experiments to minimize air scattering, enhance temperature regulation, and restrict vibrations to which these setups are highly sensitive. Conducting these interferometer experiments in vacuum can eliminate the need for much larger, more costly environment suppression (such as found at \NIOFh)  and help make perfect-crystal neutron interferometry experiments more robust to external environmental changes. 
The vacuum chamber used in Saggu \textit{et al.} to demonstrate improved phase stability had a \SI{305}{mm} diameter and a height of \SI{219}{mm}  (\SI{16}{L}),  which only allowed for the interferometer itself to be placed in vacuum. 
\par
The vacuum chamber \emph{Olympus} described here is 15$\times$ larger and can accommodate most if not all the necessary components for any given perfect-crystal neutron interferometer experiment or  \pendo measurement. 
Implementing environmental isolation will improve precision scattering length measurements, enable experiments utilizing cold samples, and will be necessary for future split-crystal interferometry. 
\par
A dedicated neutron interferometer cryostat has been tested at \NIOFa{} for use in a variety of neutron interferometry experiments such as mapping phase transitions in spin systems and measuring the temperature dependence of \pendo interference. 
We have demonstrated the ability to observe  contrast and phase of a sample over a range of temperatures from \SI{10}{K} up to \SI{300}{K}. 
Combined with environmental isolation from \emph{Olympus}, this opens up new frontiers in neutron interferometry to characterize magnetic domains, phase transitions, and spin properties in a wide range of materials. While the demonstration with a cryogenic sample discussed in Sec. \ref{sec:cryo:expt} did not take advantage of either the DFS interferometer or the vacuum chamber thermal isolation, a future generation of this experiment will work to incorporate multiple isolation techniques.

\section{Acknowledgments}\label{sec:acknowledgments}
This work was supported by The National Institute of Standards and Technology (NIST), the US Department of Energy, Office of Nuclear Physics, under Interagency Agreement
89243019SSC000025, the National Science Foundation (NSF) under Grant No. NSF-2209590, Triangle Universities Nuclear Laboratory (TUNL), the Natural Sciences and Engineering Council of Canada (NSERC) Discovery program, the Canadian Excellence Research Chairs (CERC) program,
the Natural Sciences and Engineering Research Council of
Canada (NSERC), and the Canada First Research Excellence
Fund (CFREF).

\bibliographystyle{elsarticle-num}
\bibliography{main_bibliography}

\end{document}